\documentclass[twocolumn,aps,preprintnumbers,prd,superscriptaddress,longbibliography,10pt]{revtex4-1}
\usepackage{amsmath,amssymb,amsfonts,mathtools}
\usepackage{epsfig}
\usepackage{graphicx}
\usepackage{slashed}
\usepackage{color}
\usepackage[normalem]{ulem}

\begin{document}

%\title{Two-Component Picture of A Light-Travelling Proton}
\title{Fundamental Properties of the Proton in Light-Front Zero Modes}

\author{Xiangdong Ji}
%\email{xdji@sjtu.edu.cn}
\affiliation{Department of Physics, University of Maryland, College Park, MD 20742, USA}
%\affiliation{Tsung-Dao Lee Institute, Shanghai Jiao Tong University, Shanghai 200240, China}

\date{\today}

\begin{abstract}
For a proton in the infinite momentum frame, its wave function contains a zero-momentum
part (light-front zero-modes) originated from the modification
of the QCD vacuum in the presence of the valence quarks, exhibiting a light-front
long-range order in the quantum state. This ``non-travelling'' component of the proton contributes to its
fundamental properties, including the mass and spin, as well as the naive-time-reversal-odd 
correlations between transverse momentum and spin. The large momentum effective theory (LaMET) provides
a theoretical approach to study the physical properties of the proton's
zero-mode ``condensate'' through lattice simulations.
\end{abstract}

\maketitle
A proton travelling at the speed of light is an idealized limit
considered by Feynman as a good approximation to the structure of high-energy protons
involved in hard collisions~\cite{Feynman:1969ej}.
While Feynman's parton model can be obtained by
boosting a bound state to the infinite momentum frame (IMF)~\cite{Weinberg:1966jm},
very quickly, theorists realized that it can be simply be obtained
by quantizing field theories using the light-front (LF) coordinates~\cite{Chang:1968bh,Kogut:1969xa},
a dynamical formalism that had been proposed years earlier by Dirac~\cite{Dirac:1949cp}.
The LF coordinates $\xi^\pm = (\xi^0\pm \xi^3)/\sqrt{2}$
have become the standard choice to understand the factorizations
of high-energy processes in quantum chromodynamics (QCD)
and the properties of the proton in the IMF~\cite{Collins:2011zzd,Sterman:1994ce}.
Since 1980's, the light-front quantization (LFQ) has been advocated as an important
approach to solve the hadron bound states in QCD~\cite{Brodsky:1997de}.
Indeed, LFQ can provide a wave function of the proton as seen by light-like probes,
possibly making the proton structure as easy to understand
as non-relativistic quantum many-body systems.

One particular attractive feature of LFQ is the kinetic distinction between
hadron states and the QCD vacuum. In the equal-time quantization,
the QCD vacuum contains complicated non-perturbative field configurations, and
the hadronic excitations are built on it from the same
degrees of freedom. On the LF, however, the QCD vacuum is made of quanta purely of zero
longitudinal momentum $k^+=0$, whereas the moving proton appears to contain just
partons $k^+\ne 0$. Thus, one can construct the proton state on the top of the perturbative
vacuum without worrying about the interference from the true
vacuum itself~\cite{Brodsky:1997de}. This can be realized
at least in a truncation scheme in which only $k^+>\epsilon$ modes
are the effective degrees of freedom, because $k^+\sim 0$ modes have
infinite light-cone energy, $k^-\sim\infty$, and shall be removed (or cut off)
from the hamiltonian in the Wilsonian approach of the renormalization
theory~\cite{Mustaki:1990im,Glazek:1993qs}.
This is convenient if one is interested in only the leading LF observables,
such as the twist-two parton distribution functions (PDFs).
For subleading ones, the effects of the zero-modes cannot be ignored.

In this paper, we advocate that when the proton travels at the speed of light,
its state contains, in additional to finite-$x(=k^+/P^+)$ partons,
zero-momentum quark-gluon modes differing from the QCD
vacuum. This zero-mode component of the proton, like
superfluids in condensed matter physics, has a long-range order in the
light-front $\xi^-$ direction and therefore behave like
a ``condensate''. The light-cone condensate contributes to the proton's
fundamental properties, including its scalar charge, mass, sub-leading LF components of
spin and momentum, and generates single-spin asymmetries in high-energy scattering.
However, the LFQ formalism generally has difficult to deal with the zero-modes directly,
except in certain off-light-cone regularizations~\cite{Lenz:1991sa,Naus:1997zg,Yamawaki:1998cy}.
On the other hand, the large momentum effective theory (LaMET) proposed
recently~\cite{Ji:2013dva,Ji:2014gla} provides an approach to study the properties
of this component through matching the finite-momentum Euclidean calculations
to the formalism of LFQ.

In LF theory, all particles and states must have longitudinal momentum $k^+\ge 0$. Since
the vacuum has momentum $0$, it contains particles all with $k^+=0$.
Schematically, one might
express the QCD vacuum wave function in a Fock expansion,
\begin{eqnarray}
         |{\bf 0}\rangle&&=\sum_{n=0}^\infty \int d^2\vec{k}_{1\perp}...d^2\vec{k}_{n\perp}\phi(\vec{k}_{1\perp}, ...\vec{k}_{n\perp}) \nonumber \\
   &&\times          \delta^2(\vec{k}_{1\perp}+...+\vec{k}_{n\perp}) a^\dagger_{0,\vec{k}_{1\perp}}...a^\dagger_{0,\vec{k}_{n\perp}}|0)
\end{eqnarray}
where $|0)$ is the perturbative vacuum, and all creation operators couple to overall-flavor and -color singlet.
[This expression is understood in light-cone gauge
$A^+=0$. The details, however, still depend on the choice of the boundary condition for the guage field $A^i(\xi^-,\vec{\xi}_\perp)$ at $\xi^-=\pm \infty$] When spontaneous symmetry breaking happens,
the vacuum expectation values of the order parameters become non-zero and the contribution
comes entirely from the vacuum zero modes.
In fact, it has been shown that in many cases that by taking a careful limit of a regulated LFQ,
the vacuum condensates are recovered from the zero-modes~\cite{Zhitnitsky:1985um,Lenz:1991sa,Burkardt:1995eb}.
In the case of the chiral condensate $\langle {\bf 0}|\overline{\psi}\psi|{\bf 0}\rangle$, the zero-mode
contributions implies long-range order along the light-cone direction $\xi^-$,
\begin{equation}
   \langle {\bf 0}|\overline{\psi}(0))W(0,\lambda n)\psi(\lambda n)|{\bf 0}\rangle \xrightarrow{\lambda\to\infty} \chi (=\langle {\bf 0}|\overline{\psi}\psi|{\bf 0}\rangle)\ ,
\end{equation}
where $n^\mu$ is a vector along $\xi^-$ direction,
$\lambda$ is the light-cone distance, and $W$ is a Wilson-line gauge link.

Consider now the proton state with LF momentum $P^+$ in the LFQ (assuming for convenience
zero-residual momentum: $P^+=P^-=M/\sqrt{2}$, with normalization $\langle P|P\rangle =
(P^+/M)\delta(0^+)\delta^2(\vec{0}_\perp)$).
The sum of all
quata in a Fock state, must satisfy
\begin{equation}
        \sum_i k_i^+ = P^+
\end{equation}
Some of these will have $k^+_i=0$, just like the quanta in the vacuum. The separation of $k^+=0$ quanta
from those of $k^+\ne 0$ might be done with momentum discretization~\cite{Maskawa:1975ky,Pauli:1985ps}, however,
the consequences we consider here are independent of the technical details.
We write schematically the proton state as,
\begin{eqnarray}
         |P\rangle&&=\sum_{m,n}^\infty \int \Pi_{i=1}^n
          [dx_i d^2\vec{k}_{i\perp}]\Pi_{i=1}^m  [d^2\vec{k}_{i\perp}] \nonumber \\
   &&\times        \delta(x_1+...+x_n-1) \delta^2(\vec{k}_{1\perp}+...+\vec{k}_{n+m\perp}) \nonumber \\ && \phi(x_1,\vec{k}_{1\perp}, ...,0,\vec{k}_{n+m\perp}) a^\dagger_{x_1,\vec{k}_{1\perp}}...a^\dagger_{x_n,\vec{k}_{n\perp}}|{\bf 0}_{nm}\rangle
\end{eqnarray}
where the quanta with zero $k^+$ have been included in zero-modes $|{\bf 0}_{nm}\rangle$,
and the ones with non-zero $x_i=k_i^+/P^+$ ($0<x_i<1$) have been shown explicitly.
It shall be warned, however, that the color, flavor, spin, and total transverse momentum of
$|{\bf 0}_{nm}\rangle$ need not vanish.
Even for those $|{\bf 0}_{nm}\rangle$ with the vacuum quantum numbers, It is likely that none of them is the same as the
true vacuum zero-modes $|{\bf 0}\rangle$. All quanta are coupled to the quantum numbers of the proton.

Therefore the light-travelling proton must has two components: The first is
the ordinary quanta of the longitudinal-momentum carriers, may be called the normal
component. The other component is the modification to the QCD vacuum,
the light-cone zero-mode component. It is tempting to call
the later component as ``condensate'' because it generates light-cone
long-range order in the proton. However,
such is more of the kinematic effect of singular Lorentz transformation,
does not signal any macroscopic occupation of single-particle states because of the transverse degrees of freedom.
The two components are quantum-mechanically entangled, and it would be
interesting to discover the nature of entanglement. While the normal component has been at the
active study for many years, no much has been said about the zero-mode component in the literature.

We now consider a number of examples, showing the
physical effects of the proton's light-cone zero modes.

The first is the scalar charge of the proton, corresponding
to the matrix element of $\sigma_q = \langle P|\overline{\psi}_q\psi_q|P\rangle$, where
$q$ can be up or down quarks. Since the vacuum chiral condensate exists,
the proton matrix element represents an excess over that of the vacuum.
For a light-travelling proton, we can write down a light-cone sum rule~\cite{Jaffe:1991ra},
\begin{equation}
            \sigma_q =\int^{1}_{-1} dx e_q(x,\mu) \ ,
\end{equation}
where the twist-three light-cone distribution $e(x,\mu)$  is,
 \begin{equation}
            e_q(x,\mu) =\int \frac{d\lambda}{2\pi}e^{ix\lambda}
            \langle P|\overline{\psi}_q(0)W(0,\lambda n)\psi_q(\lambda n)|P\rangle
 \label{edef}
\end{equation}
where $\mu$ is an ultra-violet (UV) renormalization scale.
The integral over $e(x,\mu)$ may diverge at small $x$ and
a proper regularization may also be needed (e.g., an isospin triplet combination has a
better convergence). The distribution $e(x)$ likely contains a non-perturbative
$\delta(x)$ function as suggested in a model calculation~\cite{Aslan:2018tff},
\begin{equation}
          e_q(x,\mu) = \delta \sigma_q \delta(x) + e_{qn}(x,\mu) \ .
\end{equation}
Using the LF quantized quark fields, $\delta \sigma_q$ comes from the
zero-mode component of the proton wave function.
According to Eq. (\ref{edef}),
\begin{equation}
       \langle P|\overline{\psi}_q(0)W(0,\lambda n)\psi_q(\lambda n)|P\rangle \xrightarrow{\lambda\to\infty}\delta\sigma_q \ .
\end{equation}
Thus $\delta \sigma_q$ implies a long-range order
along the light cone in the proton state.

The existence of the $\delta$-function at $x=0$ for twist-three observables
has been suspected as well
for the transverse spin structure function $g_2(x)$~\cite{Jaffe:1990qh,Aslan:2018tff}.
In this case, it affects the light-cone sum rule of the axial charge
for the proton, e.g. $g_A$, calculated in terms of the transverse component
of the axial current, $\overline{\psi}\gamma_\perp\gamma_5\psi$.
Besides $e(x)$ and $g_2(x)$, there is a light-cone sum rule for the tensor charge
related to twist-three longitudinally-polarized distribution $h_L(x)$,
which contains a potential $\delta(x)$~\cite{Jaffe:1991ra,Aslan:2018tff}.
In terms of the dispersion relations related to the so-called light-cone
current algebra, the existence of the $\delta$-function contributions corresponds to a
subtraction constant, called $J=0$ fixed pole~\cite{Zee:1971em,Corrigan:1971nb,Broadhurst:1973fr}.
Such contributions exist in general for higher-twist LF observables, and
the example of $F_L(x)$ considered in the old literature
(see also~\cite{Ji:1993ey}) corresponds to a twist-four distribution.

An interesting case of the twist-four sum rules is the proton mass. In the rest frame
of the proton, its mass structure has been analyzed in terms of the
QCD hamiltonian~\cite{Ji:1994av}. What does the mass structure
look like in the IFM? We have to consider the hamiltonian in LFQ,
$H = P^-= \int d^3\xi T^{+-}(\xi)$, where
$T^{+-}$ is a component of the energy-momentum tensor, and is a twist-four operator.
The LF energy equation is,
\begin{equation}
     M = \sqrt{2}\langle P| P^-|P\rangle/\langle P|P\rangle = 2\langle P|T^{+-}|P\rangle \ ,
\end{equation}
and the mass structure follow from the decomposition of $T^{+-}$.

The renormalized QCD energy momentum tensor can be written
as a sum of the traceless and trace
parts, respectively,
\begin{equation}
     T^{\mu\nu} =  \overline{ T}^{\mu\nu} + \hat T^{\mu\nu}
\end{equation}
where the traceless part can be decomposed as
$\overline{T}^{\mu\nu}=\overline{T}^{\mu\nu}_q+\overline{T}^{\mu\nu}_g $
with
\begin{align}
\overline{T}^{\mu\nu}_q=&\overline{\psi}\gamma^{(\mu}i\overleftrightarrow{D}^{\nu)}\psi-\frac{1}{4}g^{\mu\nu}\overline{\psi}m\psi\,,\nonumber\\
\overline{T}^{\mu\nu}_g=&\frac{1}{4}g^{\mu\nu}F^2-F^{\mu\alpha}{F^{\nu}}_{\alpha} \ ,\label{tmuqg}
\end{align}
And the trace part is
\begin{equation}
    \hat T^{\mu\nu}= \frac{1}{4} g^{\mu\nu}\left[(1+\gamma_m) \overline{\psi}m\psi + \frac{\beta(g)}{2g} F^{\mu\nu}F_{\mu\nu}\right]
\end{equation}
where the anomlous dimension $\gamma_m$ will be ignored for convenience, and $\beta(g)$ is the QCD $\beta$-function.
One can then write done a LF decomposition for the proton mass,
\begin{eqnarray}
  M &=& 2\langle P|\left[\overline{T}_q^{+-}+ \hat T^{+-}_m) +\overline{T}_g^{+-} + \hat T^{+-}_a \right]|P\rangle \nonumber \\
    &=& M_q^{\rm LF} + M_g^{\rm LF} +  M_a^{\rm LF}
\end{eqnarray}
where
\begin{eqnarray}
  M_q^{\rm LF} &=& (a+b) M/2  \\
  M_g^{\rm LF} &=& (1-a)M/2  \\
  M_a^{\rm LF} &=& (1-b)M/2,
\end{eqnarray}
respectively, with $a$ is the fraction of the proton momentum
carried by quarks $a=\sum_q \int dx x q(x)$, and $b=\sum_q \sigma_q (m_q/M)$.
In the massless limit, the quark and gluon kinetic motions contribute the fractions of the proton mass
, $a/4$ and $(1-a)/4$, respectively, and the anomaly gives 1/2.

One can split the fields in each of the energy operators along the light-cone and define
the following twist-four distributions,
 \begin{eqnarray}
            Q_q(x,\mu^2) &=&\int \frac{d\lambda}{2\pi}e^{ix\lambda}
            \langle P|\overline\psi_q(0)W(0,\lambda n)(\gamma^+i\overleftrightarrow{D}^- \nonumber \\ && + \gamma^-i\overleftrightarrow{D}^+)\psi_q(\lambda n)|P\rangle  \ , \nonumber   \\
            E(x,\mu^2) &=& \int \frac{d\lambda}{2\pi}e^{ix\lambda}
            \langle P|[F^{+-}(0)W(0,\lambda n)F^{+-}(\lambda n)|P\rangle\ ,  \nonumber   \\
                        B(x,\mu^2) &=& \int \frac{d\lambda}{2\pi}e^{ix\lambda}
            \langle P|F^{12}(0)W(0,\lambda n)F^{12}(\lambda n)|P\rangle \ ,   \nonumber  \\
            A(x,\mu^2) &=& \sum_i \int \frac{d\lambda}{2\pi}e^{ix\lambda}
            \langle P|F^{+i}(0)W(0,\lambda n)F^{-i}(\lambda n)\nonumber \\ && +F^{-i}(0)W(0,\lambda n)F^{+i}(\lambda n)|P\rangle
\end{eqnarray}
where $i=1,2$.
Then one can write done the light-cone sum rules,
\begin{eqnarray}
          M_q^{\rm LF} &=& \int^1_{-1} dx \sum_q Q_q(x,\mu^2) ) \ ,    \\
          M_g^{\rm LF} &=& \int^1_{-1} dx  [B(x)+E(x)] \ , \\
          M_a^{\rm LF} &=& \int^1_{-1} dx \frac{\beta(g)}{2g} (B(x,\mu^2)-E(x,\mu^2)-A(x,\mu^2)) \ .   \nonumber
\end{eqnarray}
It is likely that $Q(x)$, $E(x)$, $B(x)$, and $A(x)$ all have
$\delta$-function contribution at $x=0$, just like the case of $e(x)$.
If so, one can express the proton mass,
\begin{equation}
           M = \delta M_0 +  M_n
\end{equation}
where $M_n$ is the contribution from the normal component of the proton,
and $\delta  M_0$ is the zero-mode contributions from the twist-four
$Q(x)$, $E(x)$, $B(x)$, and $A(x)$. In LFQ without zero-modes,
$\delta M_0$ may be regarded as an additive renormalization constant.
The proton mass or hadron mass spectra can be
computed only when the zero modes are manifestly taken into account.

Apart from the $\delta$-function contributions to the sum rules, there are also physical
observables directly sensitive to the zero-modes. An example is the single-spin asymmetry involving
a transversely-polarized proton with spin vector $\vec{S}_\perp$. One can define,
in this case, transverse-momentum-dependent (TMD) distribution, $f^{\rm TMD}$,
which has an angular correlation between
the transverse momentum of a quark and the transverse spin of the proton~\cite{Sivers:1989cc,Brodsky:2002cx}
\begin{equation}
     f^{\rm TMD} (x,k_\perp, S_\perp) \sim
     (S_\perp \times \vec{k}_\perp)^z f_{1T}^\perp(x, k_\perp)+ ...
\end{equation}
where $\vec{k}_\perp$ is the transverse momentum of a quark, and the renormalization
and rapidity scales have been omitted. The TMD distribution is related to the LF correlation,
 \begin{align}\label{eq:beam}
& f(x,\vec{k}_\perp) = \int\frac{d\lambda}{2\pi}\frac{d^2\vec{b}_\perp}{(2\pi)^2}
 e^{-i\lambda x+i\vec{k}_\perp \cdot \vec{b}_\perp}\nonumber \\
& \times \langle P| \bar \psi(\lambda n +\vec{b}_\perp)\gamma^+{\cal W}_n(\lambda n+\vec{b}_\perp)\psi(0)|P\rangle  \ ,
 \end{align}
where ${\cal W}_n(\lambda n+\vec{b}_\perp)$ is the staple-shaped gauge link,
\begin{align}
&{\cal W}_n(\xi)=W^{\dagger}_{n}(\xi)W_{\perp}^\dagger(\xi_\perp) W_\perp(0)W_{n}(0) \label{eq:staplen} \ , \\
%&W_{n}(\xi)= {\rm P}e^{-ig\int_{0}^{\infty} d\lambda n\cdot A(\xi+\lambda n)} \ ,
&W_{n}(\xi)= {\rm P}\exp\left[-ig\int_{0}^{\infty} d\lambda n\cdot A(\xi+\lambda n)\right] \ ,
\end{align}
The $W_\perp$ is a transverse gauge link at $\xi^-=\infty$,
\begin{equation}
W_\perp(\vec{\xi}_\perp) =
P\exp\left[ig\int^\infty_0 d\vec{\eta}_\perp\cdot \vec{A}_\perp(\xi^-=\infty,\vec{\xi}_\perp+\vec{\eta}_\perp)\right]  \ .
\end{equation}
In light-cone gauge, $A^+=0$ and $W_n=1$, only the gauge link at $\xi^-=\pm \infty$ survives.
Therefore, the correlation $f_{1T}^\perp$ exists only if there is a
non-vanishing $A_\perp(\xi^-=\infty,\xi_\perp)\neq 0$, which involves
zero momentum gluons $a^+(0,\vec{k}_\perp)$ or $a(0,\vec{k}_\perp)$~\cite{Ji:2002aa,Belitsky:2002sm}. They have non-zero
matrix element between states $\{|{\bf 0}_{nm}\rangle\}$ which differ by
a zero-mode gluon. Since the correlation effect can be measured
experimentally, without $\{|{\bf 0}_{nm}\rangle\}$, one cannot come up an interpretation
in LFQ formalism. Similarly, in the collinear expansion formalism of
the single spin asymmetry, the zero-mode quarks and gluons in the proton
are also a key to understand the non-vanishing twist-three
correlation functions~\cite{Qiu:1991pp}.

While all leading-twist observables such as momentum, helicity, and transversity
PDFs receive contributions only from the normal component of the proton,
all other observables are in principle sensitive to the proton's
zero modes. In the sum rule cases, the presence of the zero-model
component is necessary to restore the Lorentz symmetry, such
as in the $g_2(x)$ and $h_L(x)$ structure functions~\cite{Jaffe:1990qh,Aslan:2018tff}.
When $k^+=0$ modes are excluded from the Hilbert
space, the QCD vacuum and the proton zero-mode component
reduces to the perturbative vacuum,
\begin{equation}
           |{\bf 0}\rangle \sim |{\bf 0}_{nm}\rangle \sim |0) \ .
\end{equation}
This is the rationale behind the common statement that
the QCD vacuum is ``trivial''~\cite{Brodsky:1997de}. Being an effective theory,
QCD in LFQ with truncation must be matched to the full theory
to recover, for instance, the vacuum condensate,
in the sense that $\langle {\bf 0}|\overline{\psi}\psi|{\bf 0}\rangle
=(0|\overline{\psi}\psi|0)+ C$, where $C$ is a matching constant.
Likewise, if the proton wave function in LFQ misses
the zero-mode component, its physics can only be recovered from
matching the truncated theory to the full theory in which the
zero modes are taken into account explicitly (as in simple models~\cite{Burkardt:1992sz}), 
or be included in phenomenological parameters~\cite{Xu:2019xhk}. 

%Experimentally, probing $x=0$ partons at leading twist usually
%requires infinite energy, as the minimum $x$ in probed in an experiment
%is usually $\Lambda_{\rm QCD}/\sqrt{s}$, where $s$ in the center-of-mass energy.
%Fortunately, the single transverse-spin asymmetry can be measured
%involving the interaction of the normal Fock component with
%the zero-mode, which in fact represents an effect of quantum
%interference.

In the remainder of the paper, I argue that
the proton's zero-mode physics can be studied from
the matching conditions between the LFQ formalism
and full-QCD calculations, such as lattice field theory,
through large-momentum effective theory
(LaMET)~\cite{Ji:2013dva,Ji:2014gla}. The principle
of LaMET is that the light-front soft and collinear physics
can be learnt through physical quantities depending on 
large-momentum external states (as oppose to LF quantum-field 
correlators in the usual LFQ formalism), which can be matched to
LF observables through $1/P^2$ expansion.

For the Sivers function, the LaMET formulation
and calculations have been explored extensively in recent
works~\cite{Ji:2014hxa,Ji:2018hvs,Ebert:2018gzl,Ebert:2019okf,Ji:2019sxk,Ebert:2019tvc,Ji:2019ewn}.
Note that pre-LaMET lattice
calculations have already shown the important effects
from the Wilson-line gauge link, which
in LFQ comes entirely as the zero-mode effect~\cite{Yoon:2017qzo}.
One can also calculate the twist-three collinear matrix elements with soft
gluon or quark poles~\cite{Qiu:1991pp} using the LaMET
formalism.

One can also calculate quasi twist-three and -four distributions to
learn about the $\delta$-function contributions.
For example, for the scalar charge, one can define the Euclidean
correlation,
 \begin{equation}
            \tilde e(\lambda=zP^z, P^z) =
            \langle P^z|\overline{\psi}(0)W(0,z)\psi(z)|P^z\rangle \ ,
\end{equation}
which can be simulated on lattice, and shall be
renormalized properly to subtract the Wilson-line
self-energy contribution. One can derive a LaMET matching formula
to match  $\tilde e(x)$ to light-cone correlation function $e(x)$
in the large $P^z$ limit. A delta function contribution will show up asymptotically
as the non-vanishing correlation at large $\lambda=zP^z$. However, this
might be challenging numerically. Model studies at finite $P^z$ might help to
identify a would-be delta function as a peek at small longitudinal momentum
$x$~\cite{Aslan:2018tff}.

Similarly, LaMET can help to study the zero-mode contribution to the
proton mass in the IMF. To determine possible $\delta(x)$
functions in $Q(x)$, $E(x)$, $B(x)$, $A(x)$, we can
start from the twist-four quasi distributions,
\begin{eqnarray}
            \tilde E(\lambda=zP^z, P^z) &=&
            \langle P^z|F^{03}(0)W(0,z)F^{03}(z)|P^z\rangle
%             \\
%                        \tilde B(\lambda=zP^z, P^z) &=&
%            \langle P^z|F^{12}(0)W(0,z)F^{12}(z)|P^z\rangle \\
%                        \tilde A(\lambda=zP^z, P^z) &=&
%            \langle P^z|F^{12}(0)W(0,z)F^{03}(z)|P^z\rangle
\end{eqnarray}
etc. Potential $\delta$-functions can again
be identified from the long-range correlations of the quasi-distributions
in $\lambda$. On the other hand, at finite $x$, the twist-four
light-cone distributions might be measurable experimentally
from deep-inelastic scattering and Drell-Yan process,
although this could be very challenging.

To summarize, a proton traveling at the speed of light
has two distinct components in its wave function.
The normal component is what one usually probes in the
deep-inelastic scattering. The zero-mode component will contribute
to all the high-twist quantities, including the mass, axial 
and tensor charges, etc. One particular interesting example is the momentum-spin
correlation of partons probed through the single spin asymmetry.
The usual LFQ method cannot calculate the
``condensate'' component directly. Through the
LaMET approach, it can be studied
through matrix elements of the large momentum states in lattice QCD.
This approach is in some sense similar to LF quantization
off light-cone advocated by some previous works~\cite{Lenz:1991sa,Naus:1997zg}.

{\it Acknowledgment.}---I thank F. Aslan, S. Brodsky, M. Burkardt, C. Ji, Yu Jia, Yizhuang Liu, E. Shuryak, J. Vary, Xiaonu Xiong, Feng Yuan, I. Zahed, and Yong Zhao for valuable discussions and correspondences.

\bibliography{ji_ref}
\end{document}